\pdfoutput=1
\documentclass[twocolumn,showpacs,preprintnumbers,amsmath,amssymb,floatfix]{revtex4-1}

\usepackage{graphicx}
\usepackage{amsmath,amssymb,combelow} 
\usepackage{gensymb}
\usepackage{fdsymbol}
\usepackage{epsfig}
\usepackage{dcolumn}
\usepackage{bm}
\usepackage{rotating}
\usepackage{times}
\usepackage{units}
\usepackage{hyperref}
\begin{document}

 
\title{Electric-field induced strange metal states and possible high-temperature superconductivity in hydrogenated graphitic fibers}
\author{Nadina Gheorghiu}
\email{Nadina.Gheorghiu@yahoo.com}
\affiliation{UES Inc., Dayton, OH 45432}
\affiliation{The Air Force Research Laboratory, Wright-Patterson Air Force Base, OH 45433}
\author{Charles R. Ebbing} 
\affiliation{University of Dayton Research Institute, Dayton, OH 45469}
\author{Timothy J. Haugan}
\affiliation{The Air Force Research Laboratory (AFRL), Aerospace Systems Directorate, AFRL/RQQM, Wright-Patterson AFB, OH 45433}

\date{\today}

\begin{abstract}
In this work, we have studied the effects from increasing the strength of the applied electric field 
on the charge transport of hydrogenated graphitic fibers.
Resistivity measurements were carried out for direct currents in the nA - mA range and for temperatures from 1.9 K to 300 K.
The high-temperature non-ohmic voltage-current dependence 
is well described by the nonlinear random resistor network model applied to systems that are disordered at all scales. 
The temperature-dependent resistivity shows linear, step-like transitions from insulating to metallic states as well as plateau features. 
As more current is being sourced, the fiber becomes more conductive and thus the current density goes up. 
The most interesting features is observed in high electric fields. 
As the fiber is cooled, the resistivity first decreases linearly with the temperature and
then enters a plateau region at a temperature $T \sim 260-280 $ K that is field-independent.
These observations on a system made out of carbon, hydrogen, nitrogen, and oxygen atoms 
suggest possible electric-field induced superconductivity with a high critical temperature that was
predicted from studying the role of chirality on the origin of life \cite{Salam1}.
\end{abstract}
\pacs{74.81.Bd, 75.50.Dd, 75.70.Rf, 74.50.+r, 74.81.Bd, 74.20.Mn}

\maketitle 

\section{Introduction} 

The role played by pressure on the interlayer coupling effects leading to superconductivity (SC) is 
by now acknowledged and continues to be intensively explored.
For instance, a two-band SC model based on a Ginzburg-Landau free energy with two order parameters can explain how 
chemical pressure effects can lead to the topological SC in MgB$_{2}$ \cite{Betouras}.
In turn, MgB$_{2}$ is considered both crystallographically and electronically equivalent to nonstaggered graphite (the B$^{-}$ layer) 
that has undergone a zero-gap semiconductor-SC
phase transition by large $c$-axis chemical pressure due to Mg$^{++}$ layers \cite{Baskaran1}.
The application of high mechanical pressure leads to to high-temperature superconductvity (HTS) in compounds 
containing hydrogen (H) ions (hydrides) \cite{Gorkov}.
High pressure turns the lightest element into a HTS metal \cite{Aschroft1,Babaev} as a result of the Wigner-Huntington transition \cite{Wigner}.
H research is also critical for energy storage, rocketry with H as a powerful propellant, and controlled cold fusion.

Recently, flat band SC was found in bilayer graphene twisted at the magic angle 1.1$^{0}$.
The new method - twistronics - can be extended to other 2D correlated electronic states with potential for discovering similar SC materials \cite{Carr}.
While H is the essential component in SC hydrides, controlling the oxygen (O) content in complex oxides can lead to
new functional electronic devices \cite{Santini} as well as SC in cuprates \cite{vanDalen}.
Recently, several SC features were found for the O-implanted C fibers \cite{Gheorghiu1}.
As another means for material manipulation, charge injection can lead to new material properties or new heterostructures used 
for energy storage \cite{Kuhne}.
Electric fields are a powerful tool used by the semiconductor industry for modulating in a
controlled way the charge density in the otherwise insulating metal-oxides.
A high enough electric field can lead to a high density of electron-hole pairs or an excitonic Bose-Einstein condensate and thus 
establish HTS in chalcogenate alloys \cite{Henish}.
The formation of excitonic pairs breaks chiral symmetry and leads to insulating behavior, whereas the
formation of Cooper pairs breaks local gauge symmetry and leads to SC.
When exploring electric field effects to find new transistors \cite{Standley}, 
it is important to consider the proximity of SC to metal-insulator (M-I) transitions. 
Thus, by harvesting pressure (either mechanical or chemical) or/and field effects leads to new SC materials. 

The base physical system in this study is the polyacrylonitrile-based (PAN) T300-type C fiber having an average diameter 7 $\mu$m. 
T300 C fibers are usually heat-treated up to $T = 1500$ $ \celsius$, resulting in a graphitized material with a $\sim93\%$ C content.
The C fiber is turbostratic, with volumes of parallel nearest-neighbor C layers randomly rotated such that the overall structure  
is random on the small scale and quasi-1D on the fiber's length. The nature of the disorder and 
its effect on the physical properties of C fibers is not completely understood. Small angle X-ray scattering 
shows that the C fibers are fractal objects, with their mass scaling relationship given by $M = L^{d_{H}}$, where $d_{H}$ is the Hausdorff dimension. 
The scattering intensity for PAN-derived C fibers varies as $I(S) \propto S^{2.3}$ with $S$ between 1.5 and 3 nm$^{-1}$, 
where $S = \Delta k/2\pi$ and $k$ is the wavevector. 
The significantly lower than 3 value of $S$ mirrors the disordered nature of these C fibers \cite{Dresselhaus}. 
The room-$T$ thermal diffusivity for these C fibers is the same order of magnitude as in copper, 
$D_{t} \simeq (5-6) \times 10^{-4}$m$^{2}$/s, thus three orders of magnitude larger than for an YBCO film.
In graphite, which is a semimetal, both kind of charge carriers - electrons and holes - are potentially contributing intrinsic charges. 
Pristine T300 C fibers are characterized by a strongly electronic($n$-type) conduction.
At $T = 300$ K, a small input current (few $\mu$A) results in a $\rho \simeq 1.8$ m$\Omega\cdot$cm resistivity, 
vs. $\simeq 1.7$ $\mu\Omega \cdot$cm for copper.
Light-weight (mass density of $\rho_{m} = 1.8$ g/cm$^{2}$ after graphitization) and high-strength material, C fibers have many applications.
The tensile strength for PAN T300 C fibers is 3.6 GPa, while boron (B) C fibers composites have a Young modulus larger than 400 GPa.
C fibers are critical material for building space probes like Voyager 1\&2. 
Micro-coiled C fibers, having morphologies very similar to those of DNA, find applications such as electromagnetic wave 
absorbers or H storage materials \cite{Shen}.  The current density can be as high as $7.1 \times 10^{3}$ A/cm$^{2}$ for 
silver (Ag)-doped graphene fibers \cite{Xu} and $6 \times 10^{8}$ A/cm$^{2}$ for copper (Cu)-doped C nanotubes (CNTs) \cite{Subramaniam}, respectively. 
Hybrid C fiber-HTS materials are used to create stronger, flexible, 
and chemically stable HTS wires for SC magnets used in particle accelerators, NMR devices 
or electromagnetic interference shielding covers for spacecrafts. NbN-coated C fibers \cite{Pike} and 
YBCO-coated C fibers \cite{Pathare} have critical densities $J_{c}$ of the order of $10^{6}$ A/cm$^{2}$ and 
the upper critical fields $B_{c2}$(0) are up to 25 T in the former \cite{Dietrich}. Improving the flux density pinning 
in HTS materials is also very important \cite{Haugan} and needs to be considered for each generation of new HTS wires.
Attempts are being made for finding SC in C allotropes \cite{Ono,Pierce}.
The need for decreasing manufacturing costs as well as for guaranteeing material sources leads us to consider more abundant materials when looking for new SC materials.

In this study, we explore the effect of gradually increasing the amount of direct current on the transport properties 
of octane-intercalated PAN-derived carbon (C) fibers.
Resistivity measurements are carried out in the temperature range 1.9 - 300 K and for direct currents up to 14 mA.
The nonlinear nature of the electrical conduction in these C fibers when sourced by high
direct currents is analyzed using the Resistor Network (NRRN) model and the Dynamic Random Resistor Network (DRRN) model \cite{Nandi}, respectively.
The alkane is expected to improve the electronic transport though the free protonation (free H$^{+}$ ions) of graphite's interfaces \cite{Kawashima1}
that can even lead to SC behavior \cite{Kawashima2}.
The presence of O is also expected due to (however small) amounts of water within the fiber.
In addition, the fact that this C-H-N-O system has the same elemental composition as the amino acids of life suggests by analogy the possible observation 
of a similar physical behavior.

\section{Experiment}
Temperature-dependent resistivity $\rho(T)$ measurements  are carried out using a Gifford-McMahon cryocooler, with the vacuum controlled by 
a Laser Analytics TCR compressor and a Pfeiffer turbo pump. The temperature stabilization is done by a LakeShore 340 Temperature Controller.
Current-voltage ($I-V$) measurements are carried out using a Keithley 2430 1 kW PULSE current-source meter and a Keithley 2183A Nanovoltmeter.
Cryogenic grease \cite{LakeShore} assures good thermal contact between the C fiber and the sapphire substrate. 
Based on known values for the thermal conductivity \cite{Ekin}, it is clear that the most significant heat transfer belongs to the C fiber. 
The sample is placed on the aluminum heater block and four POGO pins \cite{Everett} are spring-pressed on the fiber.
C fibers have anisotropic $\rho(T)$-dependence, with the in-plane and out-of-plane $T$ coefficient of $\rho$ being
$\alpha_{a} \cong -3.85 \times10^{-6}/$K and $\alpha_{c} \cong 9 \times10^{-6}/$K, respectively.
The small change with $T$ of the ratio between the cross-sectional area and the length of the C fiber, $S/l$ is neglected. 
With the system immersed in a cryogenic fluid, the temperature difference between the axis and the surface of the C fiber is given 
by $\Delta T = RI^{2}/(4\pi K_{T}l)$. 
The transverse (along the $c$ axis) thermal conductivity is $K_{T} = 10^{-2}$ W$\cdot$m$^{-1}\cdot$K$^{-1}$.
When a few-mm long C fiber is sourced by $I \sim$ mA, $\Delta T \sim1$ K.
The effect of a residual resistance is eliminated by using the four-wire Van Der Pauw technique \cite{Ekin},
with the the $I$-to-$V$ gap ratio close to the required factor of four. 
The quality of the electrical Ag contacts is optically checked using an Olympus B$\times$51 microscope (inset in Fig. \ref{Fig1}).

\section{Results and Discussion} 
\begin{figure}
\includegraphics[width=3in]{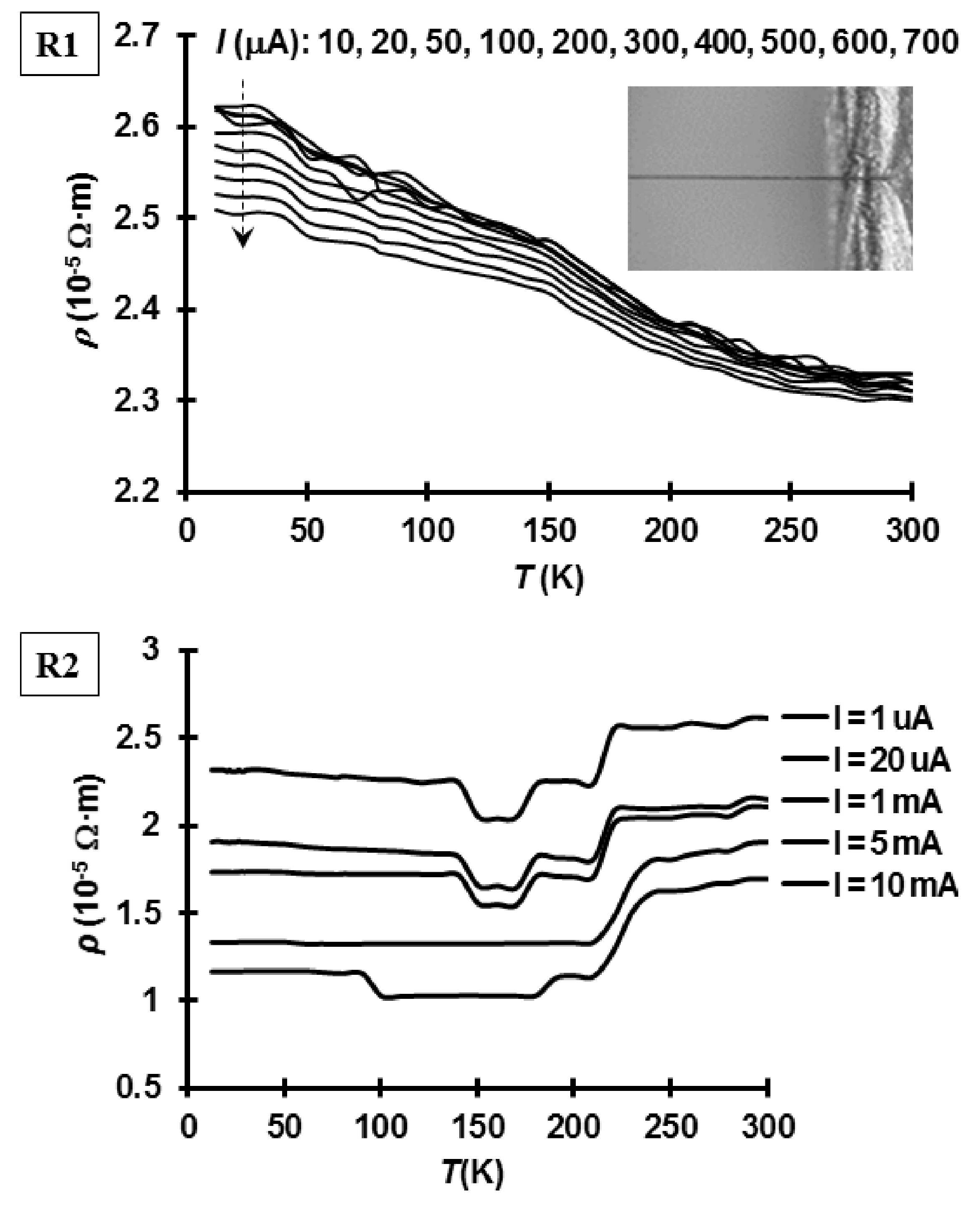}
\caption{Temperature dependence of $\rho(T)$ for two raw C fibers. Sample R$_{1}$ is more resistive vs. 
sample R$_{2}$ that shows several I-M and M-I transitions. Inset: Optical microscope image of the 7 $\mu$m in diameter C fiber 
at one of the (current) Ag contacts.}
\label{Fig1}
\end{figure}

The temperature-dependent resistivity $\rho(T)$ for raw C fibers is shown in Fig. \ref{Fig1}. While one sample (R1) is more resistive, 
another sample (R2) shows a step-like $\rho(T)$. 
The better conductivity showed by the latter raw C fiber  due to a larger amount of incorporated water content. 
I-M and M-I transitions are observed at $T \simeq $ 250, 225, 175, 150, and 100 K, respectively.
The multiple-step tunneling feature reflects the existance of abrupt changes in the electric transport (i.e., density of states) at the cooling of the sample.
The steps also point to the disordered nature of the graphitic system.
As more current is sourced through the fiber, the temperature range where steps in $\rho(T)$ occur increases.
Local maxima in $\rho(T)$ might also indicate AFM transitions.

While plateau regions in $\rho(T)$ were also found in closely-packed CNTs film formed 
by thermal decomposition on SiC and explained on the basis of the STB model \cite{Norimatsu},   
the steps between plateaus in $\rho(T)$ are quite unusual. When the control parameter is the magnetic induction $B$ instead of the temperature, 
steps in $\rho(B)$ are observed with the integer quantum Hall effect as well as 
fractional quantum Hall effect in graphite \cite{Kempa}-\cite{Kopelevich}.
In the triad $(\rho,T,B)$, both $T$ and $B$ play the role of viscosity in determining 
the local values of $\rho$. 

\begin{figure}
\includegraphics[width=3.3 in]{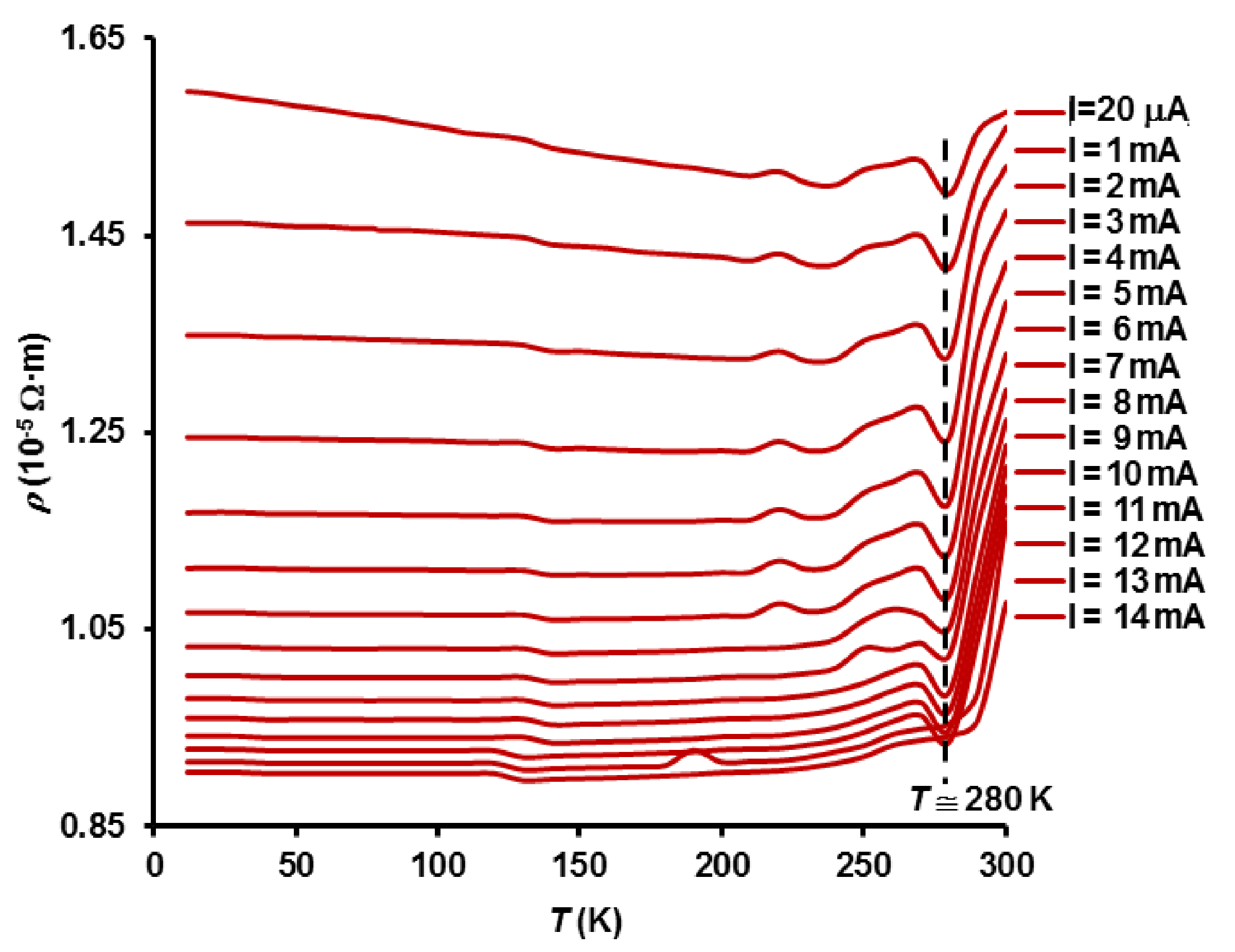}
\caption{$I$-dependence of $\rho(T)$ for the octane-intercalated C fiber.}
\label{Fig2}
\end{figure}

The $\rho(T)$ landscape for the current density measurements was significantly more richer for the octane-intercalated C fiber.
A significant drop in $\rho$ was found at $T \sim280$ K, where the raw C fiber shows only a weak minimum.
The minimum in $\rho$ at $T \sim280$ K is consistently observed for all currents (Fig. \ref{Fig2}).
The current density reaches $J_{c} = 3.6 \times 10^{4}$ A/cm$^{2}$ for $I = 14$ mA.
We attribute the difference in the response in $J$ to the H-rich octane, which can result in the hydrogenation of the $sp^{2}$ bonds 
present in the C fiber and thus improved electrical conductivity. Owning to the existence of Stone-Wales transformation defects,
it was found that diamond nanothreads hydrogenated by an organic material such as the hydrocarbon polyethylene becomes less brittle. 
Although the hydrogenated surface reduces the strength of the Van der Waals interactions, the irregular surface can compensate for the
increased interfacial shear strength between the nanothreads and the polyethylene \cite{Zhan}. Likewise, we think that the intercalation of 
the octane in between the graphite interfaces leads to the formation of hydrogenated $sp^{2}$ bonds. The C fibers become less brittle, 
this benefiting the electronic transport.
The metallization of H in the octane-intercalated C fiber is realized in the form of C-H bonds, where the $\pi$ electrons are like conduction electrons.
Then, SC-like states exists locally in complex organic molecules with conjugate bonds \cite{Kresin1,Mourachkine}.
\begin{figure}
\vspace{0.2in}
\includegraphics[width=3.3 in]{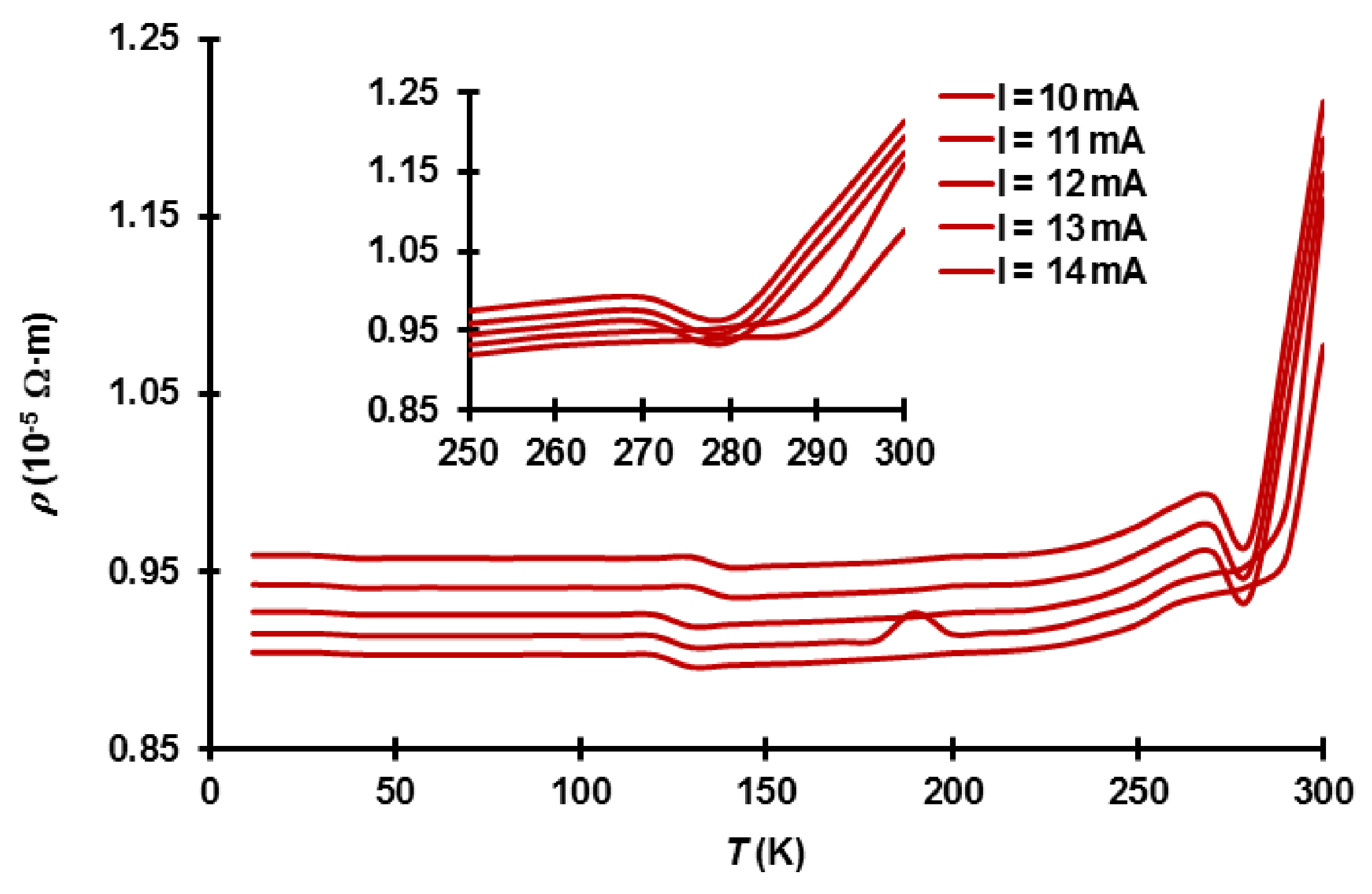}
\caption{The high-field effect (i.e., large injection currents) on the $I$-dependence of $\rho(T)$ for the octane-intercalated C fiber.}
\label{Fig3}
\end{figure}

A closer look at the $I$-dependent $\rho(T)$ curves in Fig. \ref{Fig2} reveals quite an unusual feature. 
All but two of the curves show a minimum in $\rho$ at $T \approxeq 280$ K. 
This anomaly was previously observed with PAN-derived C fibers \cite{Dejev}.
For $I < 10$ mA, $\rho(T) \sim T$ for $280$ K $< T < 290$ K and $\rho \sim T^{2}$ for $T > 290$ K.
In cuprates HTS, the $\rho \sim T^{2}$ dependence in the pseudogap regime coincides with a reduction in the electronic
specific heat to a form that is quadratic in $T$.
$\rho \sim T$ is the \textit{strange metal} (i.e., normal metal) dependence, which is attributed to the electronic system being
well described by the laws of hydrodynamics with minimal viscosity.
It results in a linear $T$-dependence of the entropy, i.e., 
$\rho$ is proportional to the electronic entropy \cite{Davison}.
The linear $\rho(T)$ dependence is characteristic to strongly correlated SCs.
Twisted bilayer graphene is a spectacular example of a strange metal 
where the strangeness might entirely result from ordinary electron-phonon scattering.
Note that the $\rho \sim T$ dependence in Dirac systems is independent of carrier density as long as the temperature is 
above the Bloch-Gr\"{u}neisen regime, where $\rho \sim T^{5}$ \cite{Hwang}.
It has even been argued that strange metals are often in the so-called Planckian limit, where the transport relaxation time is given simply 
by the temperature $\tau = \hbar/k_{B}T$. For $T = 280$ K, we find $\tau \simeq 27$ ps.
On the other hand, for $I > 10$ mA, Fig. \ref{Fig3} shows that $\rho$ first drops linearly with decreasing $T$ and then for $I = 13$ mA and $I = 14$ mA, $\rho$ almost levels out to a plateau. 
In addition, the two-step feature in $\rho(T)$ at $T \simeq 280$ K could indicate the presence of granular SC.
While a magnetic field was not applied as in \cite{Balaev}, here too we observe an abrupt change in $\rho$ at a the high-temperature, suggesting 
possible formation of SC crystallites at $T \simeq 280$ K. The following smoother transition at $T \simeq 260-270$ K might be due
to the existing boundaries between the SC crystallites.
Thus, a significant finding here is the linear $\rho(T)$ feature above $T \approxeq 280$ K that suggests a strange-metal behavior. 
Twisted bilayer graphene with its flat-energy band spectrum is a 
strange metal, where the linear $\rho(T)$ (down to $T \sim 50$ K for higher electron densities)
may be arising entirely from ordinary electron-phonon scattering 
leading to a large electron-phonon coupling constant \cite{Hwang}.
In addition, Fig. \ref{Fig3} shows also two small-amplitude transitions: a M-I-M transition at $T \sim 230$ K 
and an I-M-I transition at $T \simeq 130-140$ K.
Notice that extrapolation of the linear $\rho(T)$ below $T \sim 280$ K gives $\rho = 0$ at $T \sim 160$ K.
As mentioned before, the BCS-type mean-field critical temperature for graphene is $T_{c} = 150$ K.
The linear $\rho(T)$ extrapolation also gives $\rho(T = 230$ K) $\simeq 0.9$ $\times $ $10^{-5}$ $\Omega \cdot$ m, which is the minimum in $\rho$ 
for the plateau region. Within the Planckian dissipative transport \cite{Hwang}, where $\rho \sim mk_{B}/(n\hbar e^{2})$,
we find for the 3D charge density $  \sim 3.6 \times 10^{-19}$ cm$^{-3}$.

\begin{figure}
\center
\includegraphics[width=3.3 in]{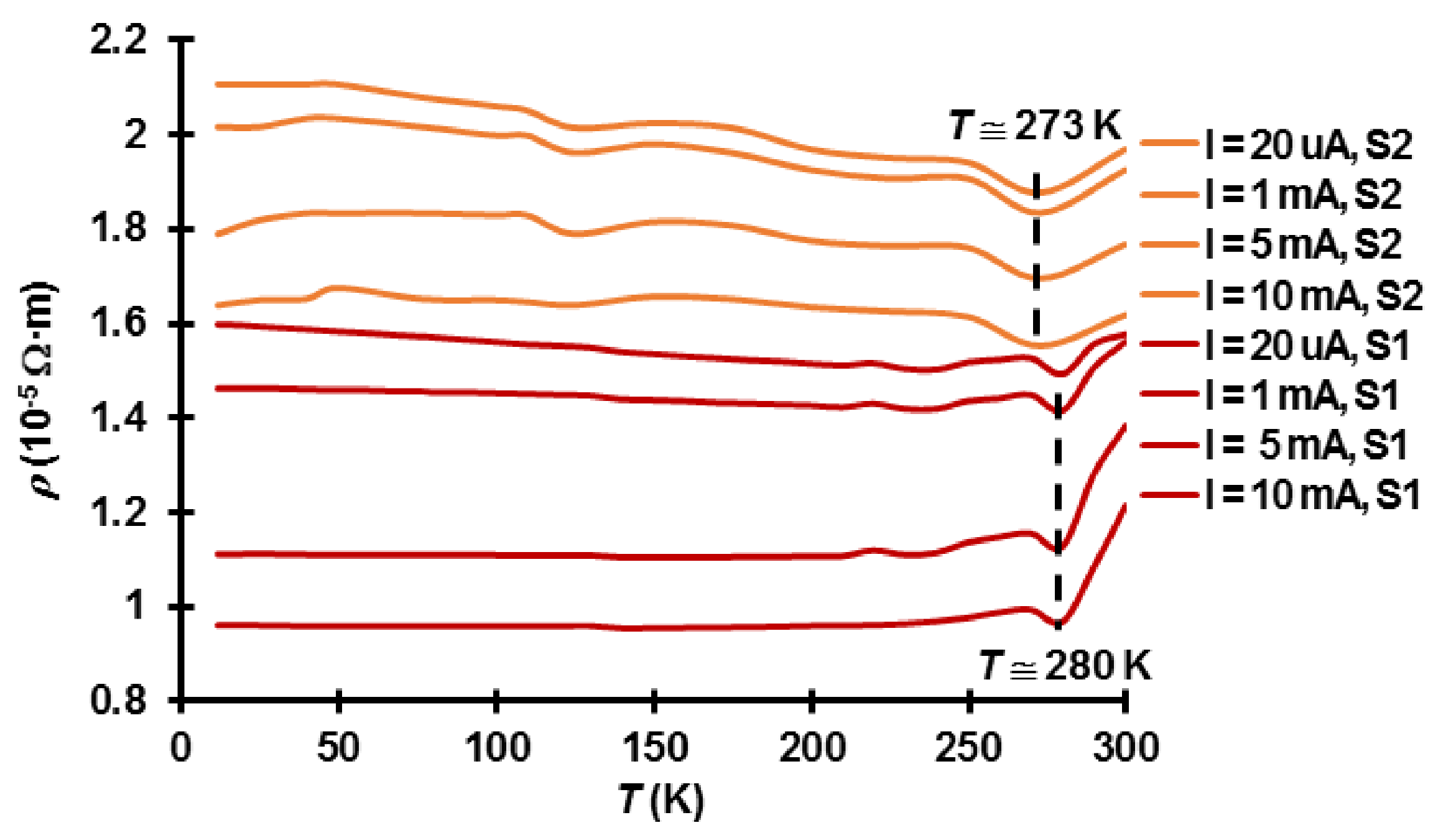}
\caption{Comparison between the $I$ dependence of $\rho(T)$ for two octane-intercalated C fibers.}
\label{Fig4}
\end{figure}

The transition at $T \sim 230$ K is quite interesting, as first-principles calculations predict HTS in metallic H \cite{Barbee}.
Correlated fluctuations between electrons and holes that result in the weakening of the Coulomb pseudopotential and band-overlap  
lead to HTS in molecular H  \cite{Johnson1}, while equal treatment of electrons and phonons result in electron-electron correlations 
that also lead to HTS \cite{Richardson}.
The high-$T$ minimum in $\rho$ was found at a lower than before, yet close $T \simeq 273$ K for another 
octane-intercalated C fiber (Fig. \ref{Fig4}).
As for the plateau feature in $\rho(T)$, we find again similarities to the Si-CNTs system mentioned before \cite{Norimatsu}.
Thus, the existence of the plateaus in both the raw and the modified sample are due to the fact that the structural features of the base system,
either raw C fiber or pristine CNTs, are invariant.

\begin{figure}
\hspace{0.3in}
\includegraphics[width=3.3in]{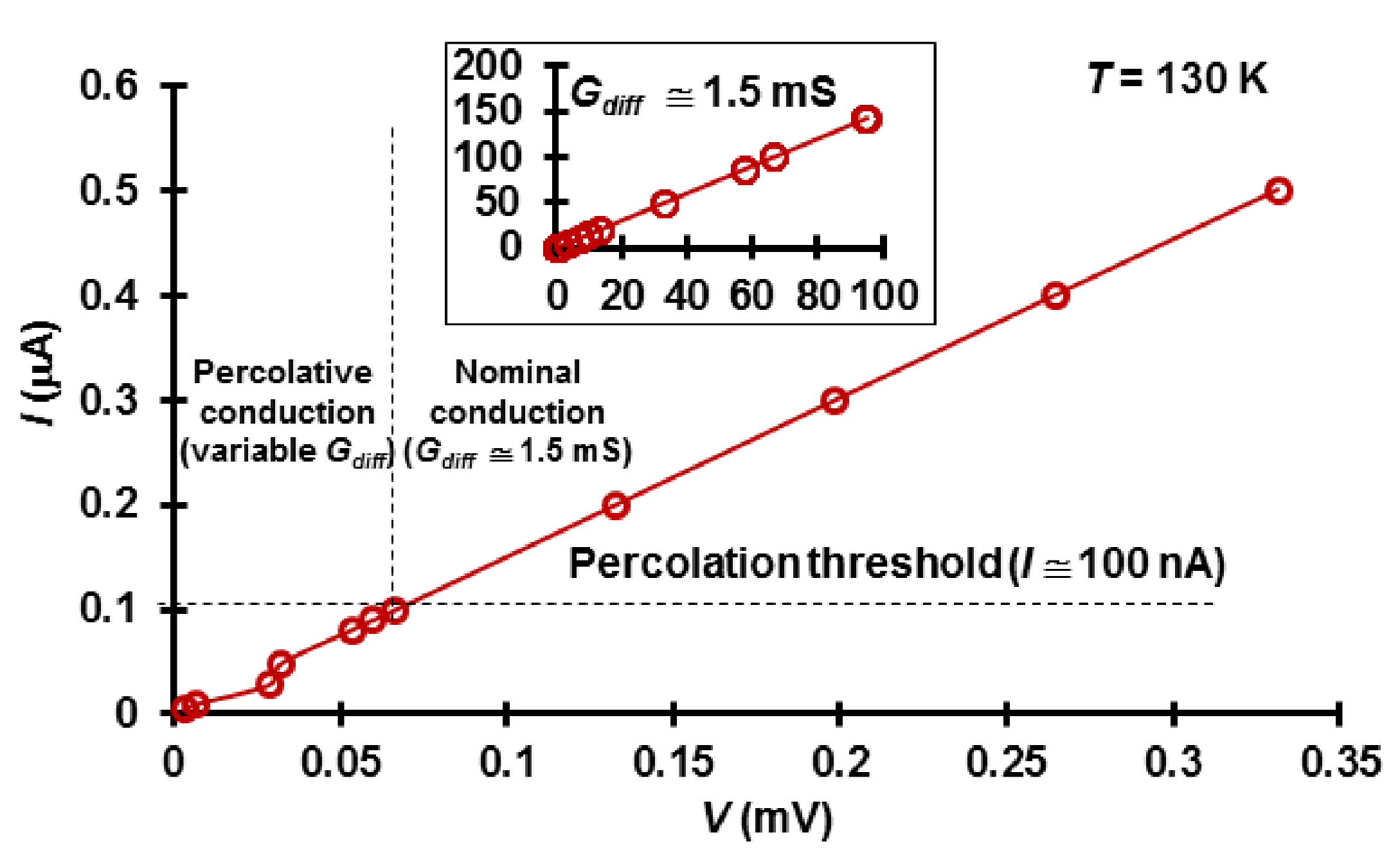}
\caption{The percolative (tunneling) nature of the electrical conduction in the octane-intercalated C fiber as revealed by 
the tuning-like behavior of the slope in $I(V)$.
The threshold current for setting the nominal regime for electrical conduction through the C fiber is 100 nA.}
\label{Fig5}
\end{figure}

The percolative nature of the electrical conduction in the octane-intercalated C fiber
is evident from the plot of the nonlocal conductance $G_{diff}$ at $T = 130$ K (Fig. \ref{Fig5}).
As the dc voltage is increased, $G_{diff}$ (i.e., the slope of $I(V)$) first starts at 1.52 mS, next drops to 0.85 mS, significantly increases to 5.80 mS, 
and finally goes back to the initial value of 1.52 mS.
The high value of 5.80 mS can be explained only by charge tunneling, which results in an increase in the number of conduction channels. 
The initially low level of the sourced current cannot sustain the feeding of initially percolated channels and the number of percolated channels then decreases.
The subsequent increase of the sourced current is enough to reestablish the initial number of conduction channels that is given by 
the electrical conductance of the sample in the nominal working regime, 
i.e., when the C fiber is sourced by not too low or not too high currents. 
During this tuning of the electrical conduction, the number of conduction channels varies from 
$\simeq 39$, to $\simeq 22$, to $\simeq 150$, and back to $\simeq 39$. The percolation threshold for the sourced current is about 100 nA.

\begin{figure}
\vspace{0.05in}
\includegraphics[width=3.3 in]{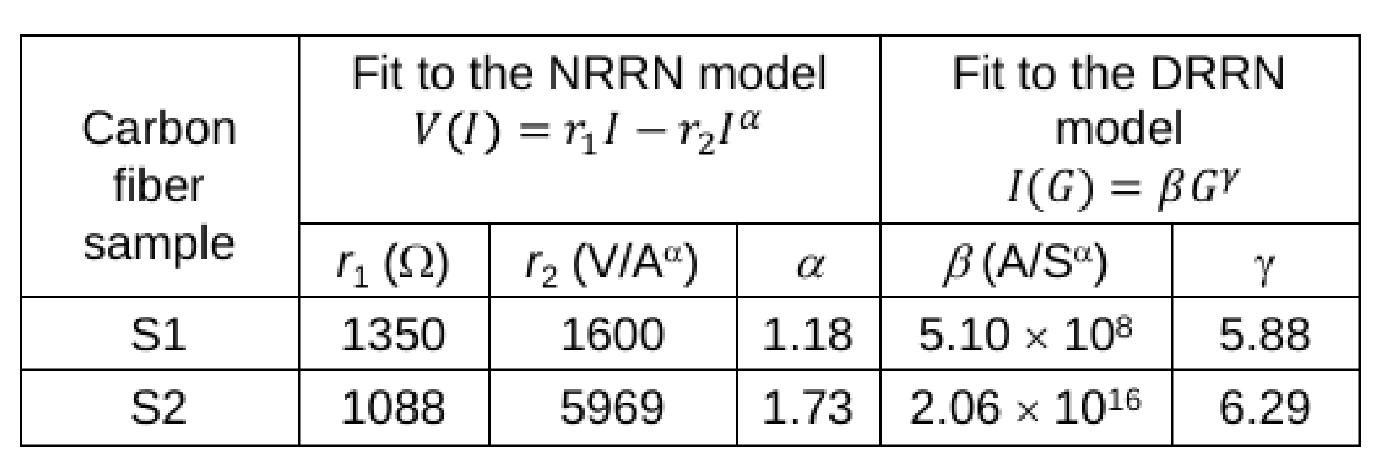}
\vspace{0.2in}
\caption{Results of fitting transport data for octane-intercalated C fibers to the NRRN model and DRRN model, respectively.}
\label{Fig6}
\end{figure}

\begin{figure}
\center
\includegraphics[width=3.3 in]{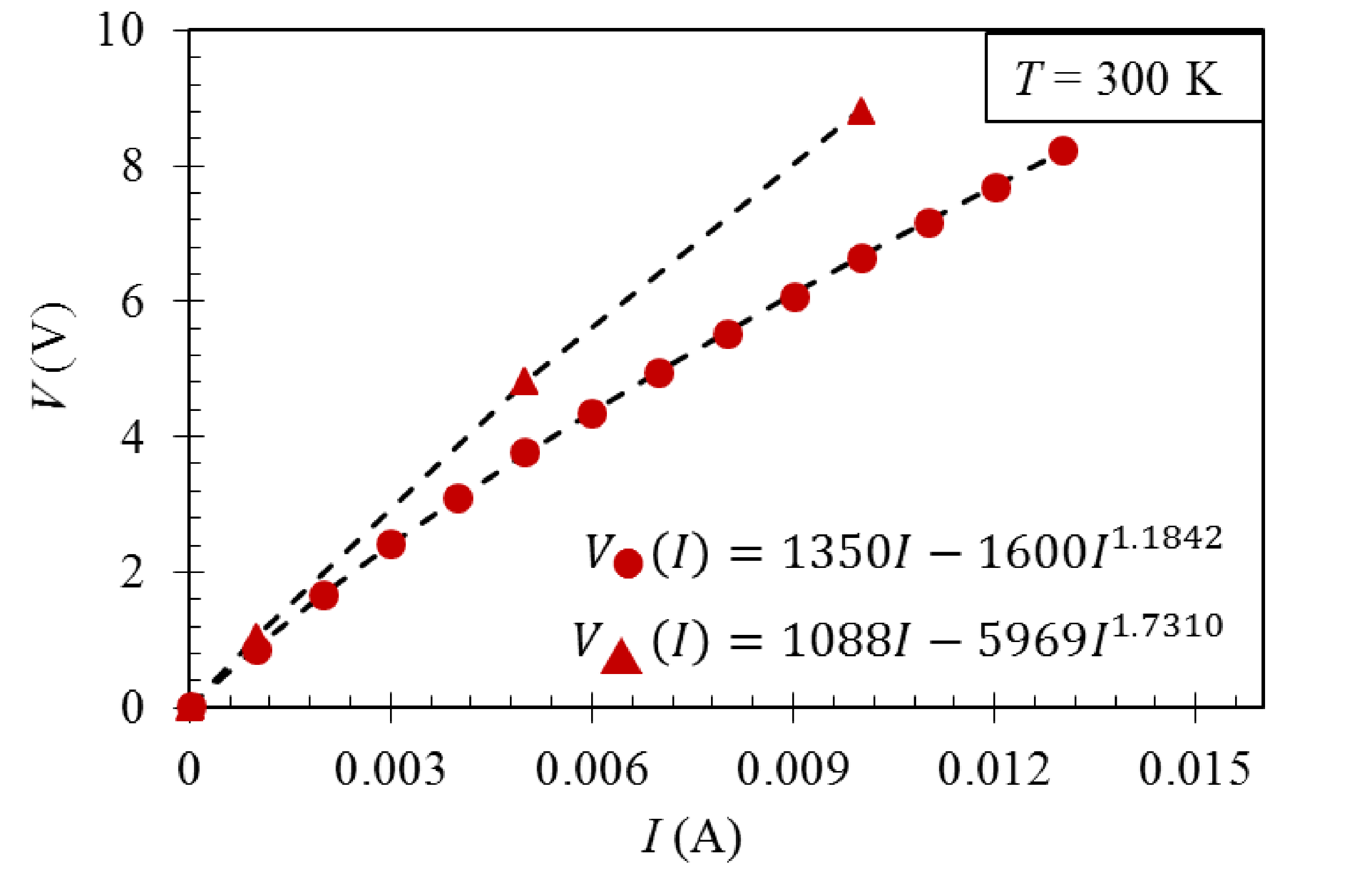}
\caption{$I(V)$ experimental data for two octane-intercalated C fibers and the fit to the polynomial
functional $V(I)$ from the NRRN model, $V(I) = r_{1}I - r_{2}I^{\alpha}$.}
\label{Fig7}
\end{figure}

\begin{figure}
\center
\includegraphics[width=3.3 in]{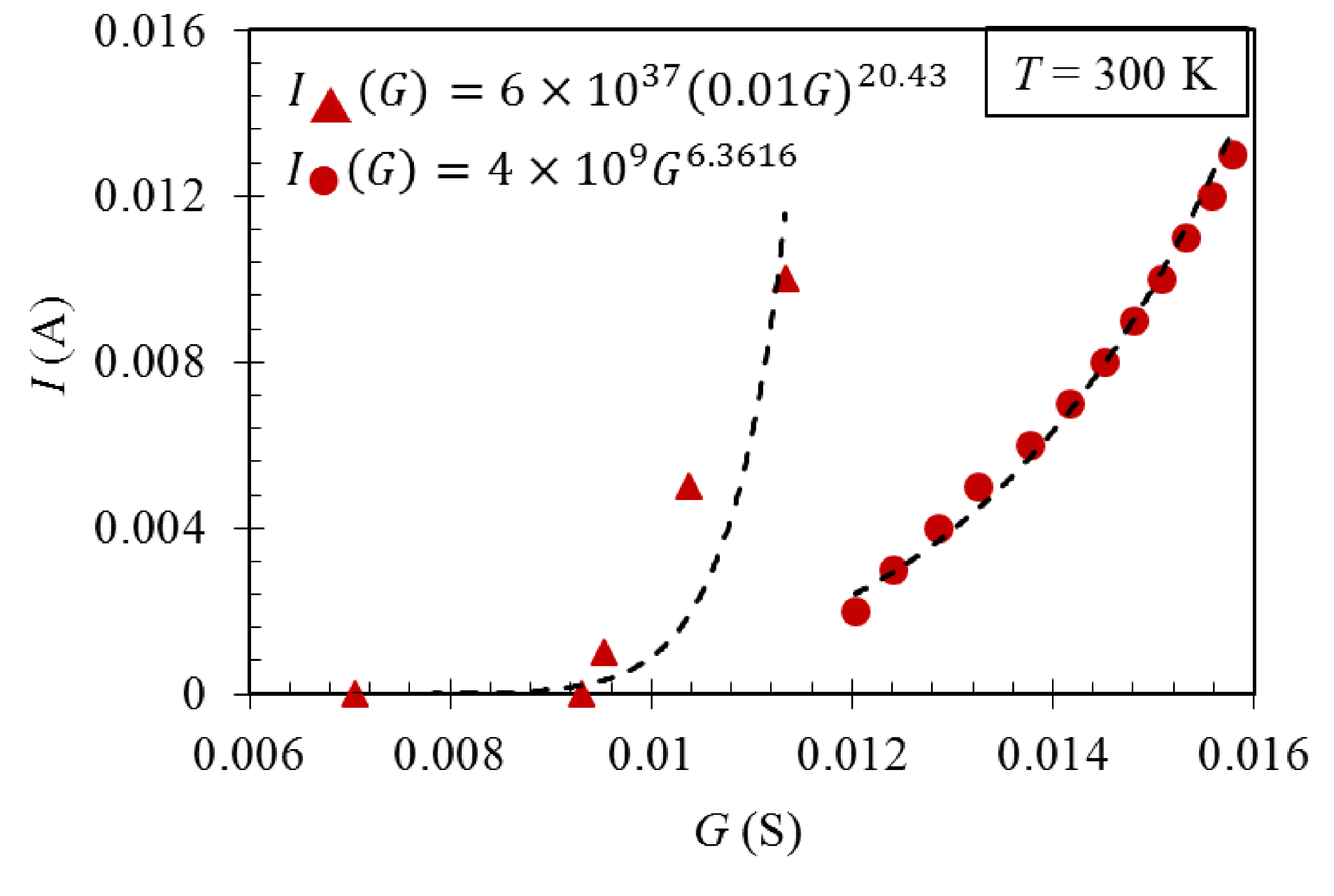}
\caption{Nonlinear current-dependent conductance experimental data for two octane-intercalated C fibers and the fit to the
power law dependence $I(G) = \beta G^{\gamma}$ from the DRRN model.}
\label{Fig8}
\end{figure}

The phenomenon known as electrical percolation is well-known to cause I-M transitions in disordered systems. The charge carriers in a semimetal,
electrons and holes, percolate via regions that have a smaller excitation gap $\Delta E$ between the carriers$\textquoteright$ energy and the edge of
either the valance band (for holes) or conduction band (for electrons) edge. Within the framework of the percolation
theory, incremental variation in the number of interconnections present in a random system suddenly causes a long-range
connectivity (clustering), resulting in the appearance of a sharp phase transition. The transition from localization
to essentially infinite connectivity is similar to a second-order phase transition that can be treated using a scaling (geometrical)
construct. Percolation is basically built into a binary system where the two components have occupied and unoccupied
bonds, resulting in the extreme values in the electrical conductance. Of great advantage to finding a solution to the
percolation model are two limiting cases: one is a M-I mixture corresponding to $G_{2}$ = 0 and the other one is a SC-normal conductor 
mixture corresponding to $G_{1} = \infty$. If $p$ is the volume fraction of either metal or
SC, then the composite$\textquoteright$s macroscopic conductance goes either from zero to a finite value or from a finite value to
an infinite value above a percolation threshold $p_{c}$ , respectively.
The common framework where the percolation phenomenon can be applied to both the I-M and normal-SC
transitions suggests the quite entrancing application of the percolation model to the C fibers in this study. 
The two models used to quantify the physical processes driving the system to a percolative I-M transition are the Nonlinear Random
Resistor Network (NRRN) model and the Dynamic Random Resistor Network (DRRN) model \cite{Nandi}. In the NRRN model, the
$I(V)$ relationship is given by: $V(I) = r_{1}I - r_{2}I^{\alpha}$, where $r_{1}$, $r_{2}$ , and $\alpha$ are fitting constants. 
In the DRRN model, there is a power law dependence $I$ on the electrical conductance (inverse of the
resistance) $G = 1/R$, $I(G) = \beta G^{\gamma}$. The major difference between the two models, NRRN and DRRN, lays in the way
nonlinearity arrives at the macroscopic scale as a result of electric conduction being networked through the microscopic
domains, of the C fiber in this case. In the NRRN model, the nonlinear character of the conduction at the macroscopic scale
is a result of the nonlinear conduction in the microscopic elements. In contrast, the DRRN model assumes that the bulk
sample can show nonlinear conduction even if the constituent microscopic elements are ohmic resistors. We have applied the
two models to the transport data obtained for two C fibers. Results are captured in the table in Fig. \ref{Fig6} and shown 
in Fig. \ref{Fig7} and Fig. \ref{Fig8} for the NRRN model (a) and the DRRN model (b), respectively. Clearly, the NRRN
model is a better fit to the experimental data for the octane-intercalated C fibers. Others found that the DRRN model better describes 
thin gold films near the percolation threshold \cite{Gefen}. The important conclusion here is that the electrical conduction is nonlinear 
at all scales of the octane-intercalated C fiber. This was actually expected, as the localization length $a^{-1}$ 
cannot possible be a match to the diameter of the C fiber. Thus, the $V(I)$
dependence for the octane-intercalated C fiber is non-ohmic and it has a $I$-dependent shape (Fig. \ref{Fig7}). 


\begin{figure}
\center
\includegraphics[width=3.3 in]{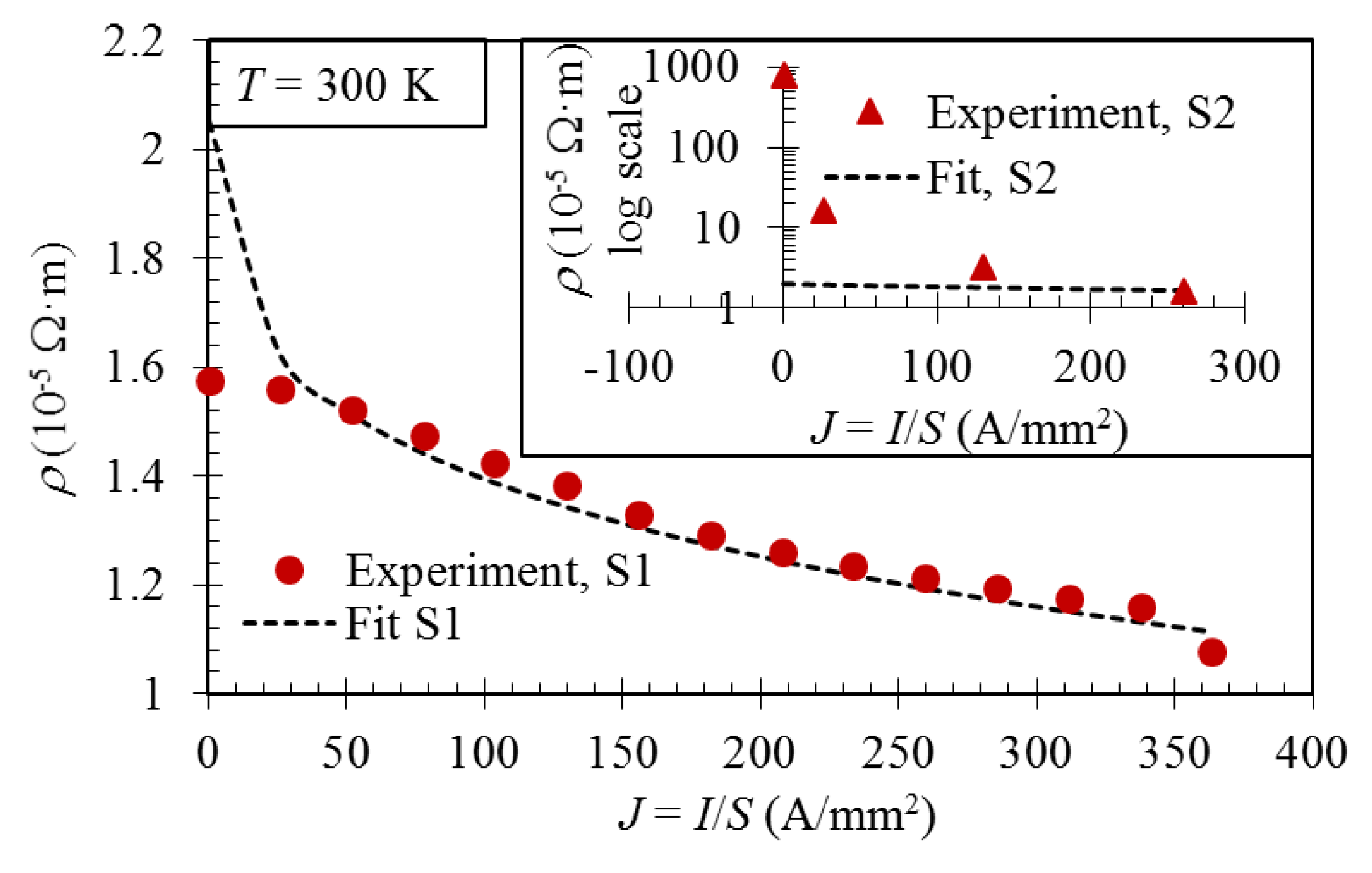}
\caption{Resistivity-current density $\rho(J)$ experimental data for two octane-intercalated C fibers at $T = 300$ K. Also shown is the fit to the dependence derived from the NRRN model, 
$\rho(J) = r_{1} - r_{2}J^{\alpha - 1}$.}
\label{Fig9}
\end{figure}

One important quantity characterizing the electronic transport through wires, in this case the C fibers, is the current density $J$.  
Solving $V(I) = r_{1}I - r_{2}I^{\alpha}$ for $I$, we find that $V = 0$ (except for $I = 0$) for $I = 0.4$ A and $I = 0.1$ A for the
two octane-intercalated C fibers denoted here by S1 and S2, respectively. The observed nonlinear electrical conduction
results in decreasing $\rho$ with increasing the sourced current density $J = I/S$, as Fig. \ref{Fig9} shows. 
For $I \ge1$ mA, there is a good agreement between experimental data for the octane-intercalated C fiber and the NRRN model, manifested in the
monotonic decrease of $\rho$ with $J$. The most likely cause for the non-ohmic character of the conduction observed with the
application of high electric fields is the field-induced tunneling. 
$\rho$ would be ideally zero or practically negligible (i.e., as for a SC) at $J_{c} \approxeq 1.0 \times 10^{6}$ A/cm$^{2}$ 
for the lower-resistance octane-intercalated C fiber (sample S1). While it is surprising to find that $J_{c}$ found
here is close to the one for HTS materials, clearly the C fiber would never reach such high $J$ due to the
extreme heating effects. Though the linear increase $\rho(T)$ above its minimum is primarily an indicative of the two-dimensional (2D) weak localization, 
a well-known linear dependence is also observed in HTS materials above their critical temperature $T_{c}$ \cite{Ruvalds}.

\begin{figure}
\center
\includegraphics[width=3.3 in]{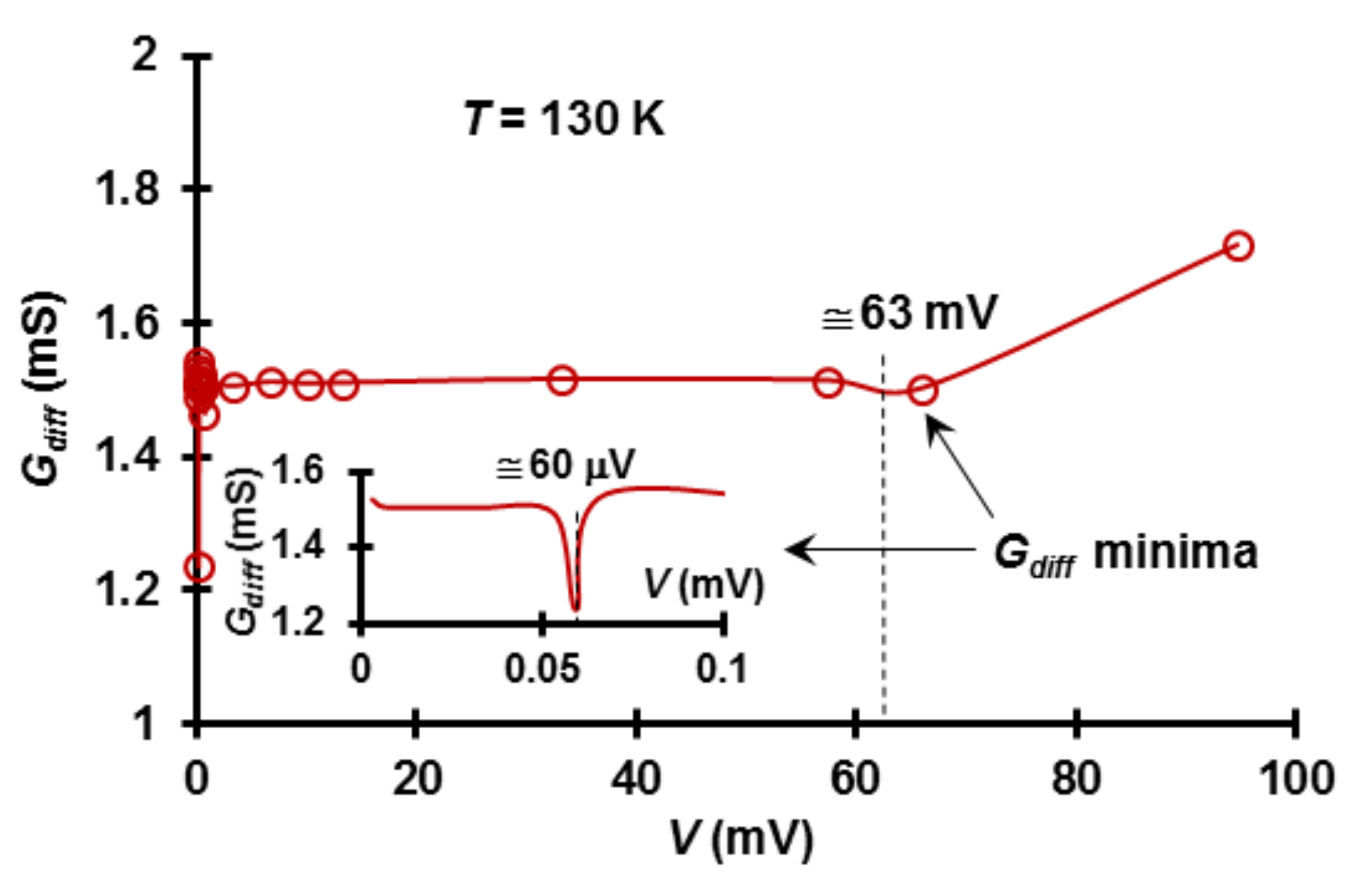}
\caption{Nonlocal differential conductance at $T = 130$ K. Two local minima are observed, with the one at the higher voltage possibly caused by 
the existence of interfacial SC currents that result in a THz wave emission.}
\label{Fig10}
\end{figure}

The $I(V)$ data for $T = 130$ K  plotted as nonlocal differential conductance $G_{diff} = dI/dV$ (Fig. \ref{Fig10}) shows two minima at voltages 
$V \approxeq 0.06$ mV (eV $\approxeq 0.096$ meV) and $V \approxeq 63$ mV (eV $\approxeq 101$ meV). 

At larger voltages, $G_{diff}$ has a monotonic increase due to percolative conduction.
As known, $G_{diff}$ shows minima at voltages $\Gamma/\left|e\right|$ ($e$ is the electron's charge), 
where $\Gamma < \Delta < G$, with $\Gamma$ the exciton superconducting
gap, $\Delta$ the superconducting gap, and $G$ the overlap band gap \cite{Bercioux}. Such minima are seen as a hallmark of the existence
of the gaped exitonic state. The second minimum here would give a BCS critical temperature 
$T_{c} \approxeq 2\Delta/k_{B} > 2\Gamma/6k_{B} \approxeq 244$ K.
The value of the gap ratio $2\Delta/k_{B}T_{c}$ considered here is at least 6, 
as previously obtained from the large gap value in \cite{Gheorghiu2}.
On the other hand, the significant drop in $\rho$ at $T \approxeq 280$ K suggests that if this would be in fact $T_{c}$, then the corresponding gap ratio would need to be 6.9.
In any case, the high value of the gap ratio clearly suggests the unconventional nature of SC in these hydrogenated C fibers.
This would also give pairing energy $\Delta_{pair} \approxeq 1.3 \times \Delta > 1.3 \times 101$ meV $\approxeq 131$ meV (or 364 K, close to temperature for 
bipolaron decay in Poly A/ Poly T DNA duplexes \cite{Lakhno}). 
Notice that for single-walled CNTs, the energy gap obtained from tunneling measurements is around $\Delta \approxeq 100$ meV \cite{Zhao}.
We also find for the exciton frequency is $f = 2\left|e\right|/h \approxeq 31 \times 10^{12}$ Hz = 31 THz, where $h \approxeq 6.62 \times 10^{-34}$ J$\cdot$s is Planck's constant.
The corresponding wavelength $\lambda = c/f \approxeq 10$ $\mu$m is of the order of the fiber's diameter $d = 7$ $\mu$m 
($c \approxeq 3 \times 10^{8}$ m/s is light speed in vacuum). Note that THz spectroscopy is used to analyze Higgs bosons, i.e., collective modes in multiband SCs like
MgB$_{2}$ \cite{Silaev} and hydrogenated graphitic fibers \cite{Gheorghiu1,Gheorghiu2}.

\begin{figure}
\center
\includegraphics[width=3.3 in]{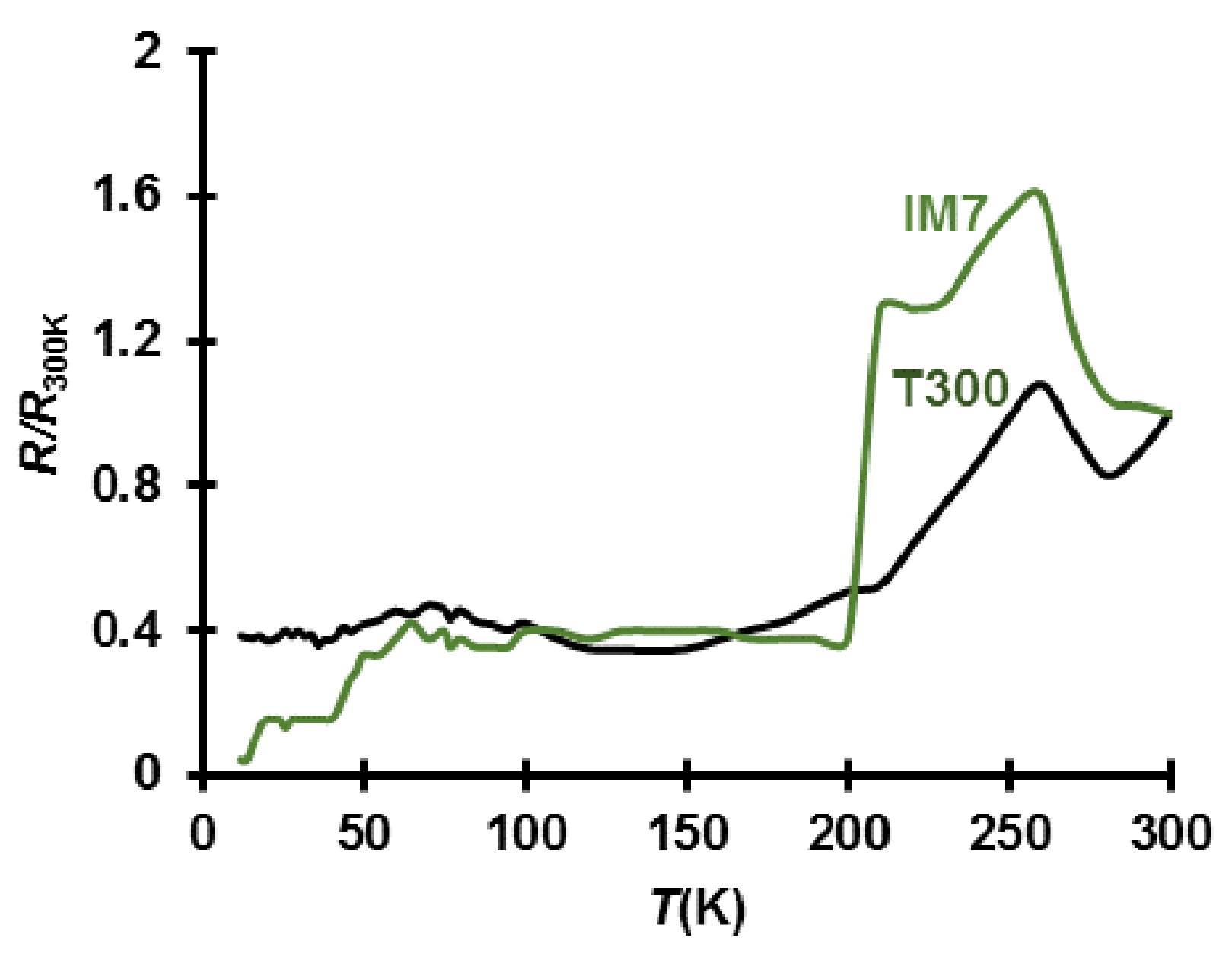}
\caption{Temperature dependence of the resistance (relative to room-temperature resistance) for bundles of C fibers T300 (N = 1000,
in black) and IM7 (N = 12000, in green) sourced by $I = 20$ $\mu$A direct current.}
\label{Fig11}
\end{figure}

I-M transitions were also observed in bundles of raw C fibers, T300
type (1000 C fibers), and IM7 type (C content 95\%) are shown in Fig. \ref{Fig11}. 
Importantly, the in-plane mean free path of electrons
becomes equal to the C fiber’s diameter when the latter is $d \leq 5$ $\mu$m, a condition that is marginally fulfilled by the IM7 C fiber
($d = 5$ $\mu$m). The resistance was plotted relative to the room-temperature value. 
There are several I-M transitions: a significant one at $T \approxeq 260$ K, close to the one found for single C fibers, at
$T \approxeq 220$ K, and at $T \approxeq 60$ K. Again, the latter occurs at the
mean-field predicted temperature for SC correlations. The IM7 also shows the better-known I-M transition at $T \approxeq 25$ K.
With all these I-M transitions, the resistance changes in an almost step-like manner. As the presence of small amount of H$_{2}$O cannot be rulled out, 
we also notice that $T = 25$ K is the ice point for CO. Recently, it was found that all comets originate from the same
point in the Universe based on the vicinity to the CO iceline in time \cite{Eistrup}. It is not unreasonable to assume that 
material properties, in particular for C allotropes, can be be traced to past physical processes in our solar system.
It was also found that the grain-surface chemistry is mainly driven by hydrogenation reactions leading to
high abundances of H$_{2}$O, CH$_{4}$, C$_{2}$H$_{6}$, and CH$_{3}$OH ices.
We also observe that the plateau feature in the $\rho(T)$ dependence is even more similar to the case of Si-CNTs system mentioned before,
when comparing the C fiber bundles here to the closely-packed CNTs \cite{Norimatsu}.
In addition, the plateau in $\rho(T)$ is centered at $T = 150$ K. 
In graphene, BCS-type SC with a mean-field
critical temperature as high as $T_{c} = 150$ K was calculated for
an electron density (2D) $n = 10^{14}$ cm$^{-2}$ and it was explained
as a Cooper-like pairing instability owning to a more efficient Coulomb screening \cite{Kveshchenko}. 
The metallic-like behavior observed with these bundles of C fibers, particularly the one
below $T \approxeq 260$ K, can be the result of parallel contributions to the electrical resistance of neighboring
interfaces (graphene planes). Significantly, granular SC might be found at the embedded interfaces \cite{Garcia,Ballestar}.
On the other hand, the I-M transition occurring at $T \sim260$ K might have a
special nature. Possible products of the heat
treatment/carbonization used in the making of C fibers are the C$_{60}$ (fullerene) clusters. Moreover, the ion-implantation
process can fragment these large C molecules. In any case, let
us assume that the C fiber contains large C molecules, with C$_{60}$
among them. At room temperature, the pure C$_{60}$ crystal adopts
a face-centered cubic structure. The molecules freely rotate
with respect to each other and orientational disorder reigns.
Then, below $T \sim260$ K, electrostatic interactions between
electron-poor ($p$-type) and electron-rich ($n$-type) regions of
neighboring molecules results in a freezing of the free
molecular rotation. In this rotation-free state, the four molecules
of the face-centered cubic unit cell become orientational non-equivalent and the C$_{60}$ undergoes a first-order phase transition
to a simple cubic crystal structure \cite{Heiney,Grivei}. 
Structural phase transition at a higher temperature precedes the charge density wave order occurring at a lower temperature in the SC state.
Thus, the structural transition at $T \approxeq 260$ K is followed by a series of steps in $\rho(T)$ with the one occurring at $T \approxeq 60$ K.
HTS in YBCO thin films and fullerenes was previously attributed to percolative conduction between clusters \cite{Szasz}. 
Metallic cluster-based SC tunneling in Josephson-linked networks, for which the presence of electronic energy
shells is similar to those in atoms and nuclei, is the key ingredient to finding room-$T$ SC \cite{Kresin2}. 
Moreover, the fullerenes can also encapsulate the H under enough pressure and the short-range order instability suggests
a two-fullerene cluster unit with the SC metallic H$_{2}$ resulting from having a H atom inside each fullerene.
Indeed, extrapolating the experimental $T_{c}(\sqrt{M})$ plot in Fig. 4 from \cite{Szasz} ($M$ is the cluster mass in atomic units 
$m_{u} \approxeq 1. \times 10^{-27}$ Kg), we find that a $T_{c} \simeq 50$ K would correspond to a mass of about $1600m_{u} = 133 m_{C}$, with 
$m_{C}$ the mass of a C atom. I.e., this could be the mass of a two-fullerene dumbbell cluster
with a H atom inside each fullerene molecule. The role of O, either implanted \cite{Gheorghiu1} or trapped 
(and under compressive stress along the crystallographic $a$-axis) \cite{Bozovic}, 
cannot be excluded when discussing the observed I-M transitions. 

\section{Conclusions}
In this work, we have studied the effects of stepping up the strength of the applied electric field on the charge transport of hydrogenated graphitic fibers.
We have found that the conduction has a percolative nature that is better described by the NRRN
model than by the DRRN model, suggesting that the fiber is nonlinear at all scales, possibly
benefiting the occurrence of SC fluctuations in this graphitic system.
The percolation model was recently applied to cuprates, where it was found that the superconducting
precursor is strongly affected by intrinsic inhomogeneity \cite{Pelc}.
The (spatial) nonhomogeneous nature of the SC gap in cuprates is reflected in the distribution of local transition temperatures and thus naturally 
leads to percolation. It might be also the case with the C fibers here, as percolation comes naturally to these fractal objects.
Percolation is also needed if assuming that SC is the primary state at ``generic'' incommensurate fillings and is being ``interrupted''
by insulating states at the commensurate fillings \cite{Chou}.
The self-organized percolative, filamentary, nature of SC in these systems might be actually captured at their room-$T$ preparation, 
thus HTS is inserted in fractal systems by design. In ferromagnetic SCs, self-organization appears to be 
the main mechanism responsible for filamentary SC by minimizing the dopant-related free energy at the formation $T$ \cite{Phillips}.
The observed electric field-induced tunneling suggests possible existence of Josephson grain-coupling in these C fibers. 
The temperature-dependent fiber's resistivity $\rho(T)$ shows several SC-like features: step-like transitions between insulating and metallic states, 
plateau regions, and high-$T$ strange metal behavior as a hallmark of non-Fermi liquid with strong electron correlations and where nematic fluctuations are important.
The interplay between disorder and SC fluctuations results in linear $\rho(T)$ at the I-M transition points.
Defects lead to weak localization, which favor electron-electron localization and hence SC correlations. 
Increased charge injection remove some of the defects and thus leads to a decrease in the electrical resistance.
The low-$T$ resistivity is plateaued to a finite value, resembling the saturated $\rho(T)$ for thin SC films that is due to either vortex 
depending or overheating effects \cite{Tamir}.
Compared to the raw graphitic fibers, the plateaus for the hydrogenated fibers are extended over larger temperature intervals 
suggesting a clearly more metallic behavior.
As known, thermal fluctuations at the SC grain boundaries lead to: a) the activation of vortices as explained by Tinkham \cite{Tinkham} 
and b) a loss of phase coherence across the Josephson junction accordingly to the model by Langer, Ambegaokar, McCumber,
and Halperin (LAMH) \cite{Langer}.
In both cases, a $2\pi$ phase slip occurs resulting to the step-like, i.e. abrupt change in temperature dependence of the electrical resistance.

We believe that this work is contributing to the growing evidence of SC located at graphite interfaces, in particular
after the samples were brought in contact with alkanes \cite{Kawashima1}-\cite{Kawashima2}, \cite{Esquinazi}.
The octane intercalation increases the carrier concentration in graphite and it was found to play a crucial role in inducing interfacial SC 
through the formation of Josephson junctions within the sample. The H-rich alkane might lead to the formation of ferrimagnetic puddles within the sample.
Possible coexistence of SC and magnetism at the interfaces of graphite that had undergone
certain modification processes, as suggested by \cite{Ginzburg}, was indeed found \cite{Gheorghiu1,Gheorghiu2}.
It was suggested that in order to reach room-$T$ SC, one must search for or artificially create
systems that experience the nontopological flat band in the bulk
or topologically protected flat bands on the surface or at the interfaces of the samples \cite{Heikkila}.
Within mean field approximations, it is shown that chiral SC domains are naturally induced by the ferromagnetic domains \cite{Tada}.
At their grain interfaces, these hydrogenated graphitic fibers might be harboring both FM-SC-FM spin valves and Josephson junctions.

The I-M transition at $T \sim 260-280$ K might be related to
\textit{chirality}, perhaps the most fundamental property in nature. 
Owning to the presence of spin carrying protons H$^{+}$ introduced by the intercalation with an alkane,
it is possible that the octane-intercalated C fiber is a chiral ferromagnetic SC.
This leads us to a much relevant connection.
The molecules of life, the DNA and RNA, are chiral and
their liquid crystal nature has been researched \cite{Stewart,Fraccia}.
Chirality can also be found in graphitic materials. 
For instance, if one rolls up a graphene sheet along the $a$-axis, a zigzag nanotube is obtained. By rolling a
graphene sheet in the direction $\theta = 30^{0}$ relative to the same axis, an armchair nanotube is obtained. 
For $0^{0} < \theta < 30^{0}$ , a nanotube called chiral will be formed. While the PAN C fiber has a turbostratic structure, one thing in common
with the DNA is the presence of elements H, N, C, and O.
Interestingly, taking into account the old ages of eucrite meteorites and their similarity to Earth$\textquoteright$s isotopic ratios 
of H, C, and N, it was demonstrated that these volatiles could have been added early to Earth, rather 
than gained during a late accretion event \cite{Sarafian}.
Even more significantly, these elemental atoms are also present in the amino acids that compose the DNA molecule.
Early studies on double-stranded DNA found that delocalized $\pi$ electrons lead to room-$T$ SC-type behavior \cite{Ladik}.
Bipolaron HTS in Poly A/ Poly T DNA duplexes was found to decay at about 350 K, which can be taken as approximate estimate of $T_{c}$ \cite{Lakhno}.
While the alkane turns the graphite from hydrophilic to hydrophobic \cite{Stando}, 
the water can still be readily adsorbed in some regions of the C fiber, thus the presence of O.
In the '90's, Salam suggested that biomolecular homochirality that originated with Pasteur can be achieved through a phase transition:
the D-amino acids would change to L-amino acids as the C$_{\alpha}$-H bond would break and the H atom would became a superconductive atom \cite{Salam1,Salam2}.
The 'D' and 'L' here stand for the right and left mirror symmetry, respectively.
Salam estimated the transition temperature to $\sim 250$ K and possibly even above 350 K.
We think that it might not be a mere coincidence with the case of octane-intercalated C fibers here, where an I-M transition occurs persistently at 
$T \sim 260-280$ K for different fibers and regardless of the amount of sourced current. Fig. \ref{Fig11} shows that the I-M transition at $T \sim 260-280$ K 
is also observed for bundles of raw C fibers. The O and the H atoms would be likely brought by any small amount of water retained 
by the C fibers during their preparation.
Real-space molecular-orbital density-wave description of Cooper pairing in conjunction with the dynamic Jahn – Teller mechanism
for HTS predicts that electron-doped water confined to the nanoscale environment of a C nanotube or biological macromolecule should superconduct
below and exhibit fast proton transport above the transition temperature, $T_{c} \approxeq 230$ K \cite{Johnson2}.
Electrocrystallization of supercooled water confined by graphene walls was also found \cite{Khusnutdinoff} at $T \approxeq 268$ K,
close to the notable temperature for the C fibers here. The linear size of the water confinement in \cite{Khusnutdinoff} is 100 \AA,
which happens to be also the length of relatively straight domains in C fibers.
Remarkable, the formation and evolution of biological structures occurred where both carbon and water were present.
More than a mere imperfection, the water retention was crucial for the formation of life. 
Rosalind Franklin and her colleagues showed that DNA formed a helix conforming to the one modeled by Watson only
within a limited yet precisely \textit{designed} range of humidity \cite{Franklin}. Is it possible that the phase transition that created the helical structures of
DNA and RNA and the one Salam refers to as superconducting and occurring around the room temperature are actually the same thing?
Worth mentioning is one particular property of hydrogen-modified graphene to behave as a shape-changing membrane \cite{Sheka} is essential to its
relation to amino acids and the SC transition at 250-300 K predicted by Salam.

Hydrates-based SC in microtubules could be responsible for quantum processing of information, such that 
the unprecedented computational power of the brain might actually come from its superconducting nature \cite{Mikheenko}.
The $T_{c} \sim 2000$ K there is close to our previously find maximum $T_{c} \sim 2360$ K \cite{Gheorghiu2}, which in turn is close to Ltlle's early 1964 prediction
for superconductivity in linear chains of organic molecules linked to certain molecular complexes \cite{Little},
also close to Schrieffer's prediction for exotic HTS \cite{Schrieffer}.
There are clear implications related to the coherent nature of the brain, its long-term memory, and human subconsciousness.
\textit{It is possible that superconductivity is organic to the living matter}. The topic needs more attention and resources, as it is 
as of paramount importance both fundamentally and for unforeseen applications.\\

\small
\centerline{\textbf{ACKNOWLEDGMENTS}}
\normalsize
This work was supported by The Air Force Office of Scientific Research
(AFOSR) LRIR \#14RQ08COR and \& LRIR \#18RQCOR100 and the Aerospace Systems Directorate (AFRL/RQ). 
We acknowledge J.P. Murphy for the cryogenics. 
Special gratitude goes from the first author to Dr. G.Y. Panasyuk for his continuous support and inspiration.

\end{document}